\documentstyle[twocolumn,floats,aps,prl,psfig]{revtex}
\begin{document}

\draft

\title{Phase Transition to Hyperon Matter in Neutron Stars}

\author{J\"urgen Schaffner-Bielich}

\address{Department of Physics, Columbia University, 538 West 120th Street, New
York, NY 10027, USA}

\author{Matthias Hanauske, Horst St\"ocker, and Walter Greiner}

\address{Institut f\"ur Theoretische Physik, J.\ W.\ Goethe-Universit\"at,
D-60054 Frankfurt am Main, Germany}

\date{\today}
\maketitle

\begin{abstract}
  Recent progress in the understanding of the high density phase of neutron
  stars advances the view that a substantial fraction of the matter consists
  of hyperons. The possible impacts of a highly attractive interaction between
  hyperons on the properties of compact stars are investigated.  We find that
  a hadronic equation of state with hyperons allows for a first order phase
  transition to hyperonic matter. The corresponding hyperon stars can have
  rather small radii of $R\approx 8$ km.
\end{abstract}


Neutron stars are an excellent observatory to probe our understanding of the
theory of strongly interacting matter at extreme densities.  The interior of
neutron stars is dense enough to allow for the appearance of new particles
with the quantum number strangeness besides the conventional nucleons and
leptons by virtue of weak equilibrium.  There is growing support that hyperons
are the first exotic particle to appear in neutron star matter at around twice
normal nuclear density \cite{Glen_book}, as recently confirmed within various
different models as effective nonrelativistic potential models
\cite{Balberg97}, the Quark-Meson Coupling Model \cite{Pal99}, extended
Relativistic Mean-Field approaches \cite{Knorren95b,SM96}, Relativistic
Hartree-Fock \cite{Huber98}, Brueckner-Hartree-Fock \cite{Baldo00,Vidana00},
and chiral effective Lagrangians \cite{Hanauske00}. The onset of hyperon
formation is controlled by the attractive hyperon-nucleon interaction as
extracted from hyperon-nucleon scattering data and hypernuclear data.  The
hyperon population rapidly increases above the critical density, eventually
even exceeding that of the nucleons.  The question arises to what extent does
the interaction between the hyperons, which is essentially unknown, influence
the overall properties of the compact star.

In this Letter, we will demonstrate that a first order phase transition to
strange hadronic matter due to highly attractive hyperon-hyperon interactions
can occur which drastically changes the global features of neutron stars
and leads to compact stars with unusually small radii of $R\approx 8$ km.
Simultaneous mass and radius measurements of a neutron star could reveal or
rule out the existence of such a novel form of matter with exotic properties,
which is in accord with our present knowledge of hadronic physics.  It is this
enormous number of hyperons in neutron stars which enables the formation of
such exotic compact stars with strangeness.

Nuclear systems with strangeness, hypernuclei, have been studied in the last
decades both experimentally and theoretically. From these studies we know that
the nucleon-$\Lambda$ interaction is attractive and that the $\Lambda$ feels a
potential of about $U_\Lambda=-28$ MeV in bulk matter \cite{Millener88}.  On
the other hand, extrapolated $\Sigma^-$ atomic data indicate that the
isoscalar potential is repulsive in the nuclear core \cite{Batty94} which is
supported by the absence of bound states in a recent $\Sigma$-hypernuclear
search \cite{Bart99}.  An attractive potential for the double strange hyperon
$\Xi$ has been extracted from the few $\Xi$ hypernuclear events \cite{Dover83}
and indirectly from final state interactions at KEK \cite{Fukuda98} and at
Brookhaven's AGS \cite{Khaustov2000}. Recently, double $\Lambda$ hypernuclear
events have been reported by E906 at AGS \cite{Ahn01} and E373 at KEK
\cite{Takahashi01} in addition to the older hypernuclear events.  The
$\Lambda\Lambda$ interaction as deduced from these double $\Lambda$
hypernuclear data is highly attractive (see \cite{FG02} and references
therein).  There is no experimental information about the other
hyperon-hyperon interactions, such as e.g.\ $\Lambda\Xi$ and $\Xi\Xi$
interactions.

A recent version of the Nijmegen soft-core potential finds extremely
attractive hyperon-hyperon interactions which even allows for the possibility
of deeply bound states of two hyperons \cite{Stoks99a} and deeply bound
hyperonic matter \cite{SG00}.  Strange hadronic matter in general will consist
of nucleons and arbitrary numbers of the hyperons $\Lambda$, $\Sigma$, $\Xi$,
and $\Omega^-$.  If the hyperon-hyperon interaction is only slightly
attractive, strange hadronic matter in bulk is bound and purely hyperonic
nuclei (MEMO's) are predicted to exist \cite{Scha93}.  The driving force is
the Pauli-blocking in the hyperon world, which forbids $\Xi$'s to decay to
$\Lambda$'s. Strange hadronic matter is metastable, i.e.\ it decays on the
timescale of the hyperon weak decay of $\tau\approx 10^{-10}$ s by loosing one
unit of strangeness.  This short-lived exotic matter can be formed in
relativistic heavy ion collisions \cite{SMS00}, as hyperons are copiously
produced in a single central event.  Neutron star matter, however, is in
$\beta$-equilibrium so that hyperon matter in neutron stars is stable on
astrophysical timescales.

In the following, we choose the standard nuclear field theory of baryons
interacting with mesons, which is solved in the mean-field approximation
\cite{Glen_book} and has been successfully applied to describe hypernuclear
data \cite{Rufa90}. The model is extended in a controlled fashion to include
the baryon octet coupled to the full nonets of scalar and vector mesons
\cite{Scha93,SM96,SG00} and is then extrapolated to large densities.  The
baryon-baryon interactions are mediated by scalar meson, $\sigma$, and vector
meson, $\omega$, and isovector meson, $\rho$, exchange.  In addition,
hyperon-hyperon interactions are modeled via hidden strange meson exchange of
a scalar, $\sigma^*$, and a vector, $\phi$, meson. The $\sigma^*$ and $\phi$
mesons couple to hyperons only.  We take the nucleon parameterization from
Glendenning and Moszkowski \cite{GM91}.  The coupling constants of the
hyperons to the $\omega$, $\rho$, $\phi$ vector mesons are fixed by using
SU(6) symmetry.  The coupling constants to the $\sigma$ meson are constrained
by the hypernuclear potential in nuclear matter of $U_\Lambda=-28$ MeV,
$U_\Sigma=+30$ MeV, $U_\Xi=-18$ MeV to be compatible with hypernuclear data
\cite{SG00}.  The remaining coupling constants of the hyperons to the
$\sigma^*$ meson are varied to investigate the effects of an enhanced
hyperon-hyperon interaction as suggested by the sparse $\Lambda\Lambda$ data.
We allow these coupling constants to scale with the number of strange quarks
of the hyperon.  The coupling constant of the $\Lambda$ hyperon to the
$\sigma^*$ meson is taken close to the corresponding nucleon $\sigma$ meson
coupling constant $g_{\sigma N}$.

\begin{figure}[t]
\psfig{file=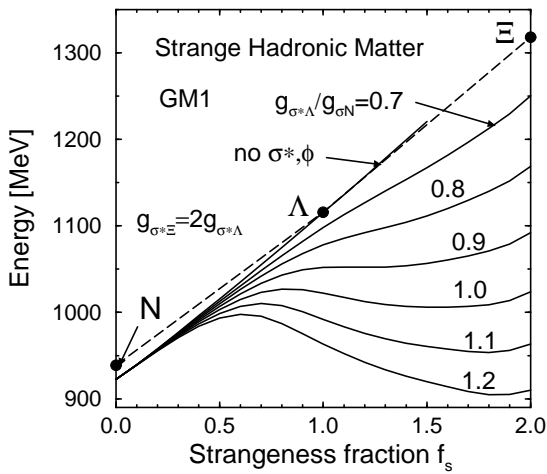,width=0.40\textwidth}
\caption{Equation of state of strange hadronic matter for different strengths
  of the hyperon-hyperon interactions. A second stable minimum appears at
  large strangeness fraction $f_s$ which can be deeper than ordinary matter.}
\label{fig:eafs}
\end{figure}

First, we discuss the stability of strange hadronic matter in bulk relevant
for heavy ion physics.  Figure~\ref{fig:eafs} shows the total energy per
baryon as a function of the strangeness fraction $f_S$, i.e.\ the number of
strange quarks per baryon.  The dashed line denotes the border between bound
and unbound strange hadronic matter. Even in the absence of the hidden strange
meson exchange, strange hadronic matter is bound up to $f_S\approx 1.5$.  If
the hyperon-hyperon interaction is taken into account, the matter gets more
deeply bound at large $f_S$. Note that the curves for $f_S\lesssim 0.4$ hardly
change and are compatible with hypernuclear data which probe at most $f_S \leq
1/3$, i.e.\ for the lightest hypernucleus ${}^3_\Lambda$H.  For
$g_{\sigma^*}/g_{\sigma N} > 0.9$, a local second minimum appears at large
$f_S>1$. This second minimum has been also seen in an effective
parameterization of the recent Nijmegen model \cite{SG00}.  Matter in this
minimum is long-lived as it can only decay into nucleons through a multiple
weak decay.  The minimum is shifted below the nucleon mass for even larger
values of $g_{\sigma^*}/g_{\sigma N}\geq 1.2$, thus creating absolutely stable
strange hadronic matter \cite{Bodmer}.  The collapse of nuclei into this
absolutely stable form is prohibited, as it would violate strangeness
conservation.

\begin{figure}[t]
\vskip-2.1cm
\centerline{\psfig{file=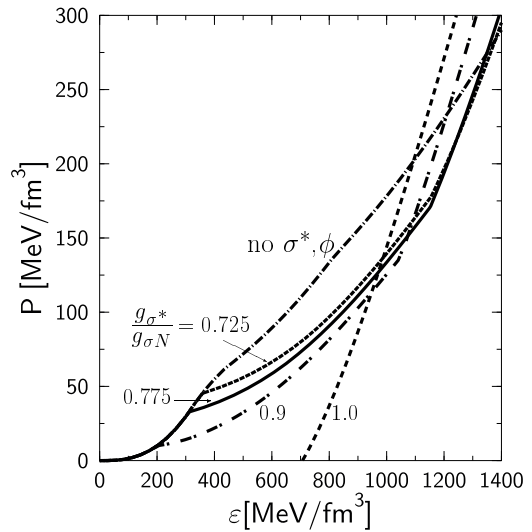,width=0.50\textwidth}}
\vskip-3.4cm
\caption{Equation of state in $\beta$ equilibrium (neutron star matter)
  for different strengths of the hyperon-hyperon interactions.}
\label{fig:eos}
\end{figure}
 
Let us discuss now the possible implications of deeply bound hyperonic matter
for compact astrophysical objects.  The equation of state (EoS) for charge
neutral $\beta$-equilibrated neutron star matter is plotted in
Fig.~\ref{fig:eos}.  A first order phase transition to hypermatter appears
which is seen as a pronounced softening of the EoS. The two kinks in the EoS
mark the beginning and the end of the mixed phase where normal and hyperonic
matter are coexisting.  The critical energy density for the onset of the mixed
phase region is lowered for stronger hyperon-hyperon interactions.  If a
second minimum is present for strange hadronic matter ($g_{\sigma^*}/g_{\sigma
N}\geq 1.0$, see Fig.~\ref{fig:eafs}), the EoS exhibits a finite value of the
energy density even for vanishing pressure, indicating that hypermatter
becomes self-bound. The corresponding compact star is then bound by the
interaction not by gravity. We stress that strange hadronic matter in the
model does not need to be absolutely stable to produce self-bound compact
stars. The presence of a second minimum, be it meta stable or absolutely
stable, seems to be sufficient to generate self-bound hyperon stars!

\begin{figure}[t]
\vskip-2.1cm
\centerline{\psfig{file=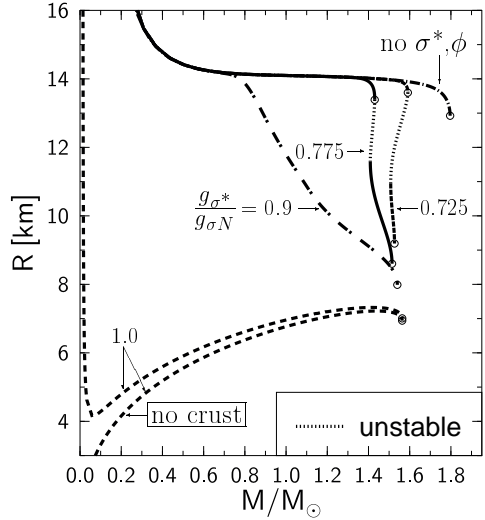,width=0.50\textwidth}}
\vskip-3.5cm
\caption{Mass-radius relation for neutron stars with a highly attractive
  hyperon-hyperon interaction.}
\label{fig:mr}
\end{figure}

The global feature of the neutron star changes drastically when the
hyperon-hyperon interaction is switched on, even for small hyperon coupling
constants (see Fig.~\ref{fig:mr}).  Without hidden strange meson exchange we
find a maximum mass of $M_{\rm max} = 1.8 M_\odot$ with a minimum radius of
$R_{\rm min} = 12.9$ km. The maximum density in the center of the star reaches
$\rho_c = 0.78$ fm$^{-3}$ which corresponds to about five times normal nuclear
density.  Increasing the hyperon-hyperon interactions results in a lower
maximum neutron star mass. A second stable solution appears for a range of
parameters ($0.78 > g_{\sigma^*}/g_{\sigma N} > 0.71$) constituting a third
family of compact stars \cite{Gerlach68}. It is located beyond white dwarfs
and ordinary neutron stars, with similar masses as predicted for neutron stars
but with considerably smaller radii.  The new solution originates from the
phase transition to hypermatter.  Compact stars belonging to this third family
contain a pure core of deeply bound hypermatter consisting of about equal
amounts of nucleons, $\Lambda$, and $\Xi$.  The central baryonic densities of
these cores are quite high, between $1.1$ fm$^{-3}< \rho_c < 2.0$ fm$^{-3}$.
The characteristic radius ranges from $8.6$ km $< R < 11.6$ km (see
Fig.~\ref{fig:mr}), which is considerably smaller than for ordinary neutron
stars. Therefore, the measurement of two neutron stars with similar masses but
distinctly different radii will serve as a unique signal for the existence of
neutron star twins.  The possibility of neutron star twins and a third family
of compact stars has been raised earlier in connection with pion condensation
and quark stars \cite{Burkhard}, and more recently for the phase transition to
strange quark matter within the MIT bag model \cite{GK2000} and perturbative
QCD \cite{FPS01}, and for kaon condensation \cite{Banik01}.  If the
hyperon-hyperon interaction is increased to $g_{\sigma^*}/g_{\sigma N} \geq
0.8$ , the two separate solutions disappear. The neutron star mass rises
continuously with energy density.  For self-bound hyperon stars
($g_{\sigma^*}/g_{\sigma N}=1.0 $), we calculate radii of $4.2$ km $< R < 7.3$
km.  Hyperon stars can have radii as small as 4.2 km for compact object with
masses as low as $M\approx 0.05M_\odot$.  The core is solely composed of
hypermatter which is surrounded by a halo of nuclei and electrons. If one
neglects the outer crust of these self-bound hyperon stars the corresponding
curve starts from the origin and an upper boundary for the radii exists of
$R<7.2$ km (see Fig.~\ref{fig:mr}, curve labeled 'no crust').  At first
glance, the mass-radius relations as discussed here are looking strikingly
similar to the ones proposed for strange (quark) stars
\cite{Haensel86,Hanauske01}. Strange stars are built of absolutely stable
strange quark matter and can have smaller radii than normal neutron stars.
Nevertheless, the maximum mass and radius for strange stars is close to the
one for an ordinary neutron star, $M_{\rm max} = 1.5-2M_\odot$ with $R\approx
10$ km when using the MIT bag model \cite{Haensel86}. Within the
Nambu--Jona-Lasinio model, these values are a little bit smaller, $M_{\rm max}
= 1.23 M_\odot$ with $R\approx 8$ km \cite{Hanauske01}.  Hyperon stars, the
hadronic counterparts of strange stars (as derived in the model used here)
have extreme nuclear properties.  They reach central baryonic densities of up
to $\rho_c = 2.1$ fm$^{-3}$ for the most massive objects, where effects from
the hadronic substructure will get important.  The region, where hadronic
equation of states are applicable, might in fact be rather small due to large
$N_c$ arguments \cite{FPS01}.  Hyperon matter can be transformed to strange
quark matter by strong interactions, as they have similar strangeness
fraction. Then, hyperon stars can form a doorway state for the formation of
strange stars as no strange quark matter seed is needed \cite{Haensel86}.

The detection of compact stars with small radii combined with small masses
($M\approx M_\odot$ or below) would signal the existence of a novel form of
matter, be it strange matter or hypermatter, which does not need to be
absolutely stable.  Recently, the radius of the isolated neutron star RX
J185635-3754 has been extracted by various groups \cite{Pons2002}.  Using the
new Chandra spectra, the radius for a black-body emitter turns out to be only
$R_{\textsc{}\infty} = 6$ km. If such a small radius is confirmed, it would
signal the existence of hypercompact stars. Nevertheless, effects of an
atmosphere can increase that value up to $R_{\textsc{}\infty} = 15\pm 3$ km
\cite{Pons2002}.

\begin{figure}[t]
\vskip-2.1cm
\centerline{\psfig{file=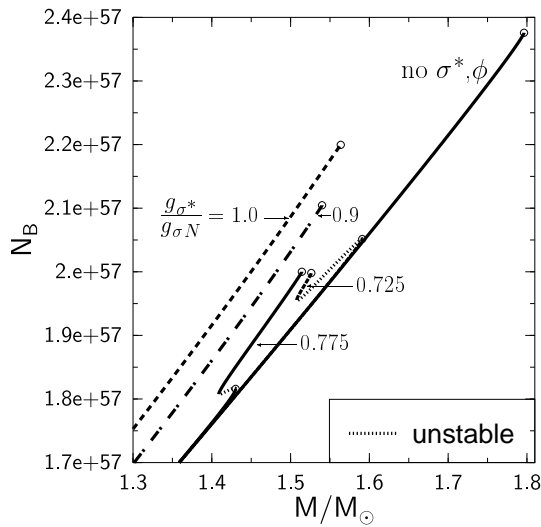,width=0.50\textwidth}}
\vskip-3.5cm
\caption{The total baryon number $N_B$ versus the gravitational mass $M/M_{\odot}$.} 
\label{fig:mbm}
\end{figure}

The conversion of a neutron star to a hyperon star should be a dynamical
process, namely a nonspherical collapse which approximately conserves the
number of baryons.  Figure~\ref{fig:mbm} depicts the baryon number as a
function of the gravitational mass of the compact stars. Note, that for a
fixed baryon number, the twin star is energetically favored compared to
ordinary neutron stars.  The mass difference from an ordinary neutron star to
its twin is about $0.03 M_{\odot}$ which corresponds to a conversion energy of
about $0.5\times 10^{53}$ erg.  Therefore, the collapse to a twin star might
have similar properties as a super- or hypernova collapse \cite{Kam83}.  The
additional release of energy due to the formation of a hyperon star in a
supernova event will generate a second energetic shock front in addition to
the standard prompt shock (similar to strange star formation as discussed in
\cite{Ben89}). The conversion to hyperon matter in compact stars will contain
an interplay of astrophysical observables, such as the spinup effect
\cite{Glen97}, the emission of gravitational waves \cite{Cheng97}, and the
emission of a $\gamma$-ray burst \cite{Olinto87}, as proposed for the
conversion to deconfined matter.  The emitted gravitational waves might be a
relevant source for LIGO, VIRGO and GEO600 \cite{Schutz98a}.

We thank Klaus Schertler and Burkhard K\"ampfer for fruitful discussions.  JSB
thanks RIKEN BNL and Brookhaven National Laboratory for their kind
hospitality.  MH thanks the Hessische Landesgraduiertenf\"orderung for their
support. This work is also supported in part by the Gesellschaft f\"ur
Schwerionenforschung, the Deutsche Forschungsgemeinschaft via
Graduiertenkolleg Theoretische und Experimentelle Schwerionenphysik, the
Bundesministerium f\"ur Bildung und Forschung and the U.S. Department of
Energy under Contract No.\ DE-FG-02-93ER-40764.


\end{document}